\documentclass[aps,twocolumn,showpacs,prb]{revtex4-1}
\usepackage[english]{babel}
\usepackage{amsmath}
\usepackage{times}
\usepackage{epsfig} 

\begin{document} 
\draft
\title{Spin and thermal conductivity in classical disordered spin chain}

\author{B. Jen\v ci\v c$^1$ and P. Prelov\v sek$^{1,2}$}
\affiliation{$^1$J.\ Stefan Institute, SI-1000 Ljubljana, Slovenia}
\affiliation{$^2$Faculty of Mathematics and Physics, University of
Ljubljana, SI-1000 Ljubljana, Slovenia}

\begin{abstract}
Transport quantities of the classical spin chain with the quenched disorder in the 
antiferromagnetic coupling $J_i$ are evaluated using the dynamical simulation at finite temperatures $T>0$ . Since the 
classical model is  nonintegrable, spin and thermal conductivities remain finite  even in the pure case. 
On the other hand, the role of disorder becomes crucial at low $T $ leading to a vanishing transport 
due to the Anderson localization within the linearized regime. The crossover from the  insulator to the conductor 
appears both for the spin and thermal transport at quite low $T^* \ll J$. Still the many-body localization
regime at $T>0$ evidenced by extremely short mean free paths can be strongly enhanced by introducing into the model 
an additional staggered field.
\end{abstract}


\maketitle

\section{Introduction}

Properties of low-dimensional spin systems have been investigated intensively since many decades. 
Since the one-dimensional (1D) $S=1/2$ antiferromagnetic (AFM) Heisenberg model  
is the first quantum many-body model considered theoretically and its ground state 
properties are exactly solvable by  the Bethe Ansatz  \cite{bethe31}, it has been the playground for 
numerous analytical and numerical approaches. In recent decades, experimental realisations of the 
spin-chain physics has been found in several classes of materials, the best and nearly ideal example
of the isotropic Heisenberg $S=1/2$ model being  cuprates  \cite{sologuben00,hlubek12}. A particular signature 
of the 1D $S=1/2$ model and its integrability are transport properties which are anomalous \cite{zotos04}.
It has been shown that the heat transport is dissipationless at any temperature $T >0$ \cite{zotos97,klumper02}, 
at the same time also the spin conductivity (diffusivity) is infinite (ballistic) in the anisotropic easy-plane regime 
(anisotropy $\Delta<1$) at any $T>0$ \cite{zotos99}.  

In contrast to clean spin chains the transport in disordered systems  is much less understood and
represents a challenge and very active  theoretical topic. It is well established that in a quantum
$S=1/2$ AFM   Heisenberg chain and more generally in the system of 1D repulsive fermions at $T=0$ any 
random quenched disorder induces localisation and  the vanishing of transport coefficients \cite{doty92,schmitteckert98}.
This is the consequence of the Anderson localisation phenomenon \cite{anderson58,dean64,kramer93} in 1D persisting 
or even enhanced in  the presence of  interaction within the ground state. 
The scenario of the many-body (quantum) localisation (MBL) \cite{basko06,oganesyan07,bandarson12} which would 
manifest itself with the   localisation and no d.c. transport at $T>0$ is investigated intensively at present,
mostly within the random Heisenberg model. So far the evidence for the latter has been found in numerical
solutions of  strongly disordered $S=1/2$ AFM  model with random local fields \cite{oganesyan07,bandarson12}, 
while some studies indicate more on a crossover into a nearly localised regime \cite{karahalios09,barisic10}. We note that there
are recently experimental realisations of disordered (but isotropic) spin $S=1/2$ chains 
BaCu$_{2}$(Si$_{1-x}$Ge$_{x})$O$_{7}$ \cite{tsukada99}, Cu(py)$_2$(Cl$_{1-x}$Br$_x$)$_2$ \cite{thede12}
and Ca-doped Sr$_2$CuO$_3$ \cite{mohan14}, where the disorder 
enters predominantly via random exchange couplings $J_i$  and has been theoretically studied 
in connection with  the concept of random singlets \cite{ma79,fisher94,herbrych13,kokalj15}.

The classical AFM  spin-chain model, which can be regarded as the $S \to \infty$ limit of the quantum
model, is  numerically much easier to deal with.  Without disorder the spin dynamics of is nonlinear and nonintegrable,
hence one obtains at $T>0$ finite transport coefficients \cite{savin05}, unlike to $S=1/2$ model \cite{zotos97} but on 
the other hand
closer to $S \geq1$ cases \cite{karada04}.  Still, the role of disorder bears some analogies to the MBL when
considering low but finite $T>0$. Namely, in the low $T$ regime one can imagine the transport being dominated 
by linear excitations which exhibit the Anderson localization in a random system \cite{dean64,kramer93}. Numerical studies 
of the classical model so far performed at $T \to \infty$ \cite{oganesyan09} did not reveal the absence of diffusion.
In the following we perform the numerical study of the d.c. spin conductivity $\sigma_s$ and the thermal conductivity 
$\kappa$ within classical spin model in the whole $T$ regime. We show that indeed at low $T \to 0$  both coefficients vanish 
due to the localization phenomena. Here the fundamental question is whether the (classical analogue) of the MBL
persists in a finite $T$ window. We find the evidence only for a insulator-conductor crossover regime, in this sense 
consistent with studies of transport in disordered anharmonic chains \cite{dhar08}. Still, the 
quasi-insulating regime characterized by extremely short spin and thermal mean free paths (MFP) $l_{s,t} \ll 1 $ 
and the crossover temperature  $T^*$ can be strongly extended by introducing a constant staggered field.  In spite of strong $T$ 
dependence, both transport MFP qualitatively satisfy the simple Wiedemann-Franz law \cite{ashcroft76} $l_s \sim l_t$.
  
The paper is organised as follows. In Sec.~II we introduce the disordered classical 1D spin-chain model. We 
define the appropriate spin and thermal currents as well as the linear-respond equations for the d.c. spin conductivity $\sigma_s$ and
the thermal conductivity $\kappa$. In Sec.~III we describe the numerical protocol for the the initial spin configuration corresponding to 
the equilibrium at chosen temperature $T$ and further simulation of the spin dynamics and calculation of d.c. transport 
properties. Results for $\sigma_s(T)$ and $\kappa(T)$ are presented in Sec.~IV for various disorder and the whole
range of $T$. A particular analysis in terms of corresponding mean free paths $l_s,l_t$, respectively, is given in 
Sec.~V and discussed in relation with the Wiedemann-Franz ratio $W=l_t/l_s$. 

\section{Model and transport quantities}

We consider in the following the 1D AFM isotropic model of classical
spins (rotors) $\mathbf{S}_{i}=S\mathbf{e}_{i}$, 
\begin{equation}
H=\sum_{i}J_{i}\mathbf{S}_{i}\cdot\mathbf{S}_{i+1}- \sum_{i}\mathbf{B}_{i}\cdot\mathbf{S}_{i},
\label{heis}
\end{equation}
where $\mathbf{e}_{i}$ are unit vectors and (exchange) couplings 
$J_{i}>0$ are antiferromagnetic on all bonds but random, in analogy with the $S=1/2$ case 
\cite{ma79,fisher94, herbrych13}.  We also  choose $J_i$ randomly distributed within an interval 
$J-\delta J\leq J_{i}\leq J+\delta J$ (with $\delta J <J$ so there is no possibility of cutting a chain)
and are as well uncorrelated between sites. 
Since we fix furtheron $S=1$, $J=1$ and the lattice parameter $a=1$ 
the only relevant parameter (in the absense of local fields, $B_i=0$) 
is the disorder strength $0<\delta J  \leq 1$.
 Our study deals with the disorder in the off-diagonal $J_i$, the motivation being also in the material realisations
where random fields  $\mathbf{B}_{i}$ are hard to justify. Nevetheless, we also comment on the influence
of the staggered field $\mathbf{B}_{i} = (-1)^i B {\bf e}_z$ which can, e.g., emerge as a mean field 
due to interchain coupling \cite{kokalj15}. We show that $B \neq 0$ has a very strong effect on
transport at $T \to 0$.  It should be also reminded that the thermal transport in the classical model with random $\mathbf{B}_{i}$
has been studied for $T\to \infty$ \cite{oganesyan09}.
 
The equations of motion for the classical model, Eq.~(\ref{heis}), are
\begin{equation}
\frac{d\mathbf{S}_{i}}{dt}= -\mathbf{S}_{i}\times\frac{\partial H}{\partial\mathbf{S}_{i}} =
-\mathbf{S}_{i}\times(J_{i-1}\mathbf{S}_{i-1}+J_{i}\mathbf{S}_{i+1}- \mathbf{B}_{i}).
 \label{eqm}
\end{equation}
Our aim is to study the spin current $j_s$ and the energy current $j_E$  which 
are defined via the continuity equations \cite{savin05,karahalios09}, 
\begin{equation}
\frac{\partial S_{i}^{z}}{\partial t}=-(j_{i+1}^{s}-j_{i}^{s}), \qquad
\frac{\partial h{}_{i}}{\partial t} =- (j_{i+1}^{E}-j_{i}^{E}), \label{curr}
\end{equation}
where $h_i$ is the local energy
\begin{equation}
h_{i}=\frac{1}{2}\left[J_{i-1}\mathbf{S}_{i-1}\cdot\mathbf{S}_{i}+J_{i}\mathbf{S}_{i}\cdot\mathbf{S}_{i+1}\right]-
\mathbf{B}_{i}\cdot\mathbf{S}_{i},
\end{equation}
The explicit expression for both currents then follow, 
\begin{eqnarray}
& j_i^{s}=J_{i} \left( S_i^x S_{i+1}^y-S_i^y S_{i+1}^x \right), \nonumber\\ 
j_i^E &= \frac{1}{2}  J_{i-1} [ J_{i-2}( \mathbf{S}_{i-2} 
\times \mathbf{S}_{i-1} ) \cdot \mathbf{S}_{i} +  
J_{i} (\mathbf{S}_{i-1}\times\mathbf{S}_{i} ) \cdot \mathbf{S}_{i+1} ]   \nonumber \\
&-\frac{1}{2}J_{i-1}\left(\mathbf{S}_{i-1}\times\mathbf{S}_{i}\right)\cdot (\mathbf{B}_{i-1}+\mathbf{B}_{i}). 
\label{je}
\end{eqnarray}
The d.c. spin conductivity $\sigma_s$ and the thermal conductivity $\kappa$
are within the (Kubo) linear response given through correlation functions \cite{kubo91},
\begin{eqnarray}
\sigma_s&=& \frac{1}{L  T}
\underset{_{\tau\rightarrow\infty}}{lim}\intop_{0}^{\tau} \left\langle 
J^{s}( t) J^{s}(0) \right \rangle dt , \nonumber \\
\kappa&=& \frac{1}{L T^2 } \underset{_{\tau\rightarrow\infty}}{lim}\intop_{0}^{\tau} 
\left\langle J^E( t) J^{E}(0) \right\rangle d t,
\label{kubo}
\end{eqnarray}
where $J^s=\sum j_i^s$ and $J^E=\sum j_i^E$ are total spin and energy currents, respectively. 
We also put $k_B=1$ so that $T \equiv T k_B/J S^2$, $\sigma_s
\equiv \sigma_s/(S a^3)$ and $\kappa \equiv \kappa/(k_B J S a^3)$ are all dimensionless. 

\section{Numerical procedure}

Thermal values of $\sigma_s(T)$ and $\kappa(T)$ for given $T$ and disorder $\delta J$
are calculated numerically  in two steps. First, we use modified 
Landau-Lifshitz-Gilbert (LLG) equations for spin dynamics \cite{garcia98}, in order to thermalise the 
system to given $T$, i.e., to  find the $T$-dependent initial spin configurations \cite{savin05},
\begin{equation}
\frac{d\mathbf{S}_{i}}{dt}=
\frac{\mathbf{S}_{i} }{ 1+\alpha^{2}}\times\left[ \mathbf{\xi}_{i}-\frac{\partial H}{\partial\mathbf{S}_{i}} - 
\alpha \mathbf{S}_{i}\times\left(\mathbf{\xi}_{i}-\frac{\partial H}{\partial\mathbf{S}_{i}} \right)\right]. 
\label{llg}
\end{equation}
with a damping parameter $\alpha$, and $\xi_{i}$ representing random 
vectors due to $T>0$ thermostat, with properties of the Gaussian
white noise,
\begin{equation}
\left\langle \xi_{i}(t)\right\rangle =0, \qquad \left\langle \xi_{i}(t_{1})\cdot\xi_{j}(t_{2})\right\rangle =
2\alpha T\delta_{ij}\delta(t_{1}-t_{2}), \label{wn}
\end{equation}
In the concrete numerical realisation we perform the calculation on
$L = 10000$ sites. The LLG thermalization, Eqs.~(\ref{llg}),(\ref{wn}),  we independently 
check via Boltzmann thermodynamic value of the energy $E(T)=\langle H \rangle$.  Clearly for $T \ll J $ longer 
thermalization times $t_0$
are  required so that we employ $t_0  \sim 1000 $.
Furtheron  final LLG spin configurations are used as initial conditions 
for dynamical spin evolution via Eq.~(\ref{eqm})  where we use the standard fourth-order Runge-Kutta 
with the small time step $\delta t \sim 0.01$ 
(tested by the conservation of spin norm $S=1$ and the total  energy $E$)
and the evolution times up to $\tau \sim 10000$.  
Kubo formulas, Eqs.~(\ref{kubo}),  are then used to 
evaluate $\sigma_s(T)$ and $\kappa(T)$ at various disorders $\delta J$.  

Such a protocol for the evaluation of transport properties, standard for generic (ergodic) systems,
should be carefully reconsidered and tested for our model, being a candidate for the MBL phenomenon.
In the case of MBL one could expect in an isolated system the lack of thermalisation,  dependence 
of the (long-time) linear response on the initial conditions etc.  Still, as shown later our results do not confirm
the strict MBL behavior  (except at very low $T$) and consistently the initial conditions do not influence
the final result. Still the closeness to MBL manifest itself in long thermalisation and evolution times $t_0, \tau \gg 1$.

\section{Results: spin and thermal conductivities}

In Fig.~1 we present numerical results for the $T$ dependence of the renormalized 
$\tilde \sigma_s=T \sigma_s$ and $\tilde \kappa= T^2 \kappa$. It follows from Eqs.~(\ref{kubo}) that such
redefined $\tilde \sigma_s,\tilde \kappa$ remain finite (and constant) at $T \to \infty$. Obtained results already reveal general features 
of the d.c. transport in pure and disordered classical chains:
\begin{figure}[ht]
\includegraphics[width=0.4\textwidth]{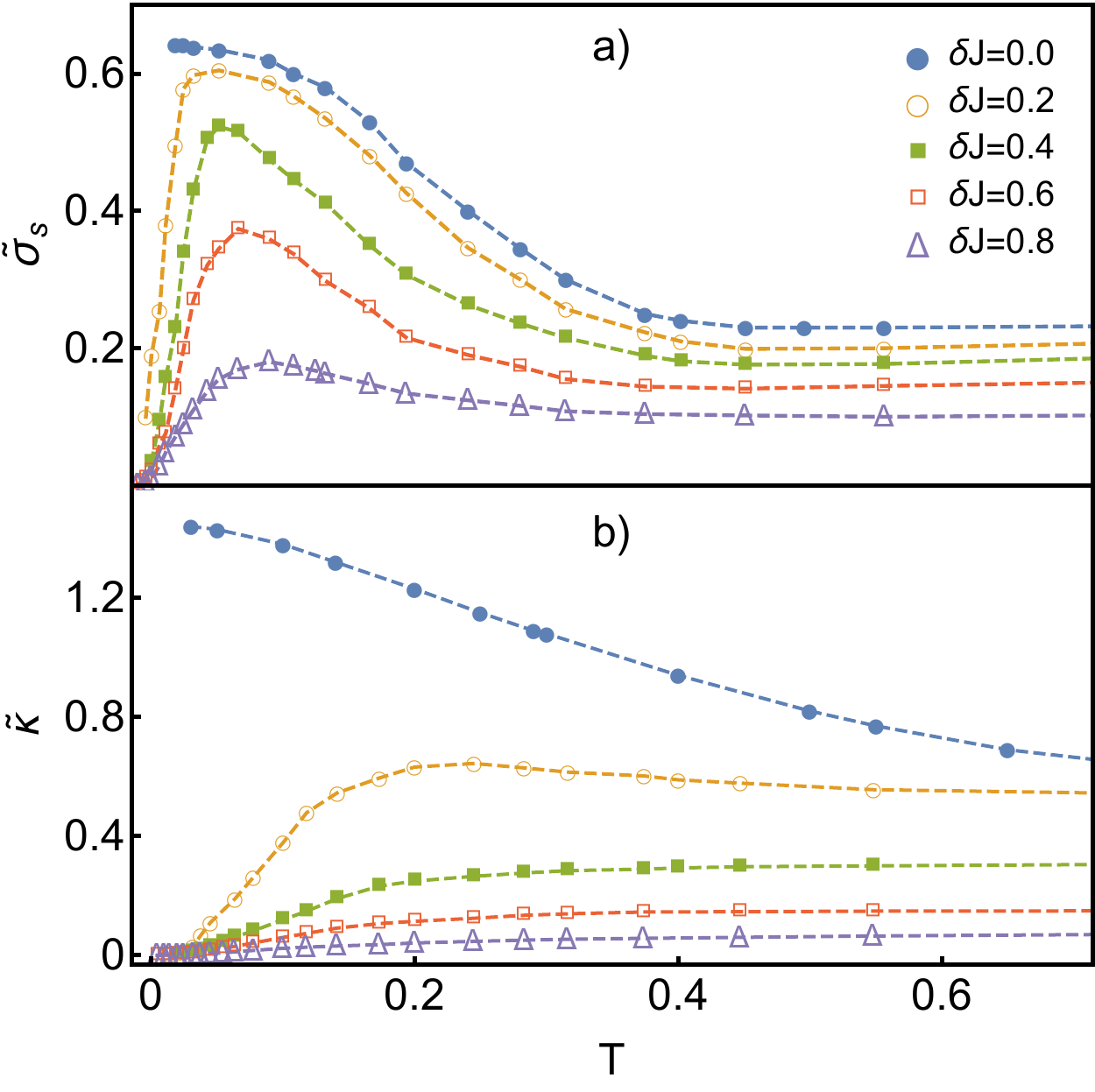}
\caption{(Color online) Temperature dependence of renormalized: a) spin conductivity $\tilde \sigma_s  =
T \sigma_s $, and b) thermal conductivity $\tilde \kappa =T^2 \kappa $ for various disorders
$\delta J=0 - 0.8 $.}
\label{fig1}
\end{figure}

\noindent a) In contrast to the quantum
$S=1/2$ spin-chain model \cite{zotos97} in the classical model both $\sigma_s(T)$ and $\kappa(T)$ are finite 
for $T>0$, even without disorder at $\delta J=0$  \cite{savin05}.

\noindent b) From Fig.~1 it is evident that $\tilde \sigma(T)$ and $\tilde \kappa(T)$ are both 
approaching constants at $T>0.5$  for any disorder $\delta J$. Our results for the clean case $\delta J=0$ for
$\tilde \kappa_\infty= \tilde \kappa (T \to \infty) \sim 0.55$ are consistent with the one previously 
obtained  \cite{savin05}.  

\noindent c) Instead of spin conductivity $\sigma_s$ more often 
considered quantity is the spin diffusivity $D_s$,+
determined by the relation   $D_s =\sigma_s/\chi_s$ 
\cite{forster75} where  $\chi_s = \langle (S^z_{tot})^2 \rangle/L T$ is 
the uniform spin susceptibility. Taking into account $\chi_s(T \to \infty) = 1/(3T)$ 
and our value from  Fig.~1a $\tilde \sigma_\infty  =\tilde \sigma_s(T \to \infty)\sim 0.25$ we would get  
the limit $\tilde D_s(T \to \infty) \sim 0.75$ 
which can be well compared with the high-$T$ expansion result  for the classical 
case $\tilde D_s(T \to \infty) = (2/5) \sqrt{10/3}=0.73 $ \cite{huber68} (note the correction
in Ref.~\cite{huber69}).

\noindent d) It is plausible that the disorder $\delta J >0$ reduces both $\tilde \kappa_\infty$ and 
$\tilde \sigma_\infty$. 
While for weaker disorder, e.g. $\delta J=0.2$, the $T\to \infty$ transport is not much effected by
the randomness, the dependence is more pronounced for $\delta J >0.4$, in particular for 
$\tilde \kappa_\infty$. The latter difference can be partly traced back  
to the explicit forms of $j^s_i$ and $j_i^E$, Eqs.~(\ref{je}),
where the exchange coupling and their disorder enters linearly in $j^s_i$,
but quadratically in $j^E_i$.

\noindent f) Both $\sigma_s(T)$ and $\kappa(T)$ are not diverging or approaching finite values,
but rather vanishing in the regime $T \to 0$ (more evident in later plots). The latter $T$ window is 
quite narrow but numerically well resolved. The absence of transport at $T \to 0$ is a direct 
indication of the localization phenomenon. Namely, Eqs.~(\ref{eqm}) can be 
linearized around the AFM ground state ${\bf S}_i^0= (-1)^i {\bf e}^z$. Resulting eigen-solutions 
of linear equations, Eq.~{\ref{eqm}), for $S_i^x$ (or $S_i^y$) are then localised  according to the Anderson localisation 
\cite{dean64,kramer93,dhar08} in 1D systems. To confirm the onset of localisation we present  in Fig.~2 
a typical example of local spin deviations  $\delta S_i^x$ obtained as the solution of the linearized Eqs.~(\ref{eqm}) 
for large disorder  $\delta J =0.8$ and  low $T = 0.02$. It is, however, evident (see later) that the validity
and feasibility of such a linear approximation is restricted to very low $T\ll 1$ at $B=0$, but can be substantially
enhanced with the staggered filed $B>0$.
\begin{figure}[ht]
\includegraphics[width=0.4\textwidth] {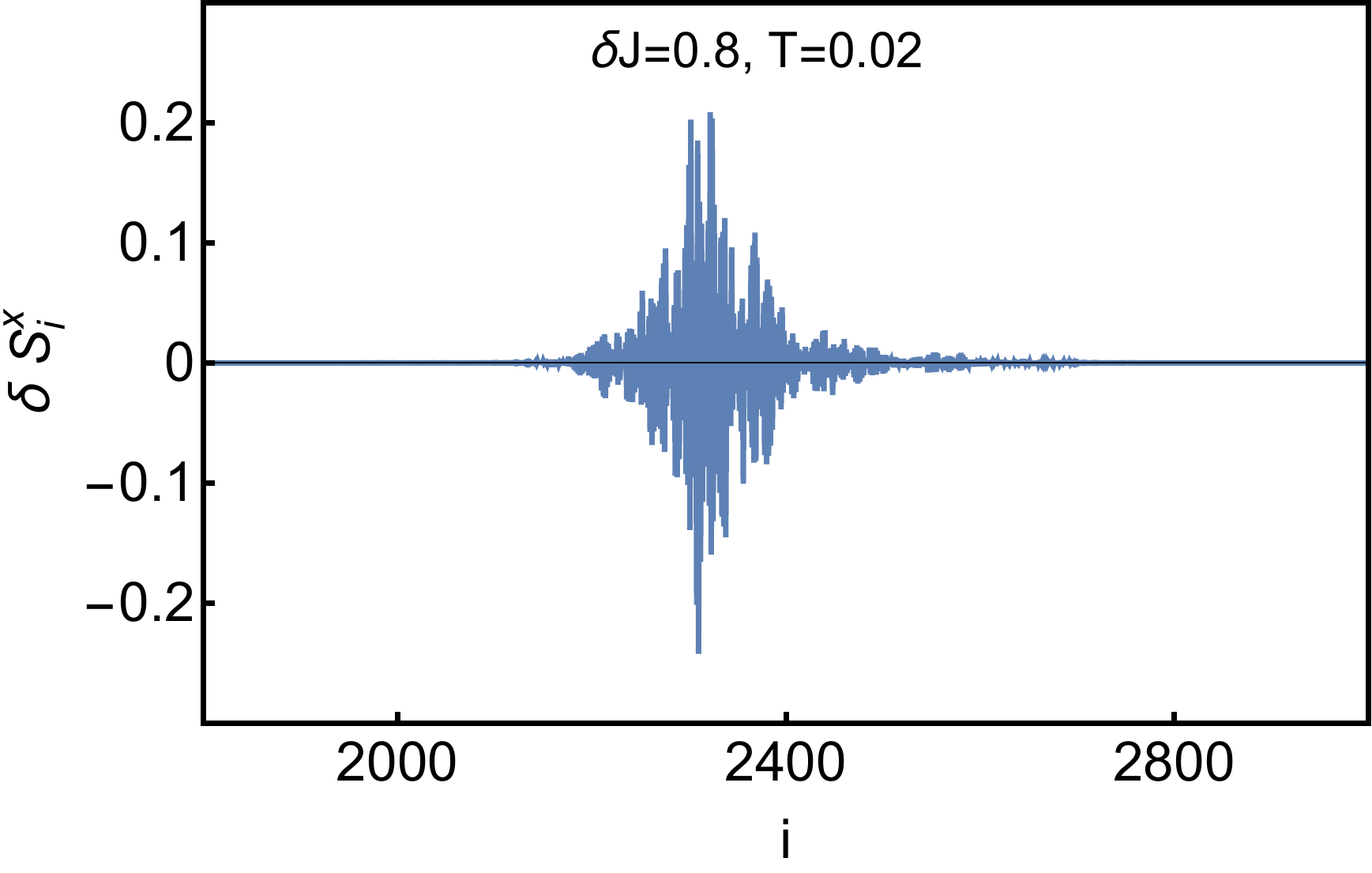}
\caption{(Color online) An example of local spin deviations emerging from the linearized 
Eqs.~(\ref{eqm}) in case of  $\delta J=0.8$ and $T \sim 0.02$.}
\label{fig2}
\end{figure}

\section{Transport mean free paths and Wiedemann-Franz ratio}

It is instructive to analyse the  transport data  (with the emphasis on low $T \ll J$ ) through 
the phenomenology of the standard kinetic theory. One can introduce the concept of thermal (heat)
MFP $l_t$ which is in our case the transport MFP of AFM magnons representing 
the relevant  low-$T$ excitations with a constant magnon velocity $v = 2J$ (emerging form the
classical AFM dispersion $\epsilon_q=  2J \sin q $). $l_t$ then enters the thermal conductivity 
as $\kappa=C_{V} v l_t$  where $C_V = (1/L ) dE/ dT$ is the specific heat of the system.
On the other hand, the spin-diffusion 
MFP $l_s$ can be most reasonably  extracted via the diffusion constant  $D_s = v l_s$.
In general both MFP are different $l_s(T) \neq l_t(T)$. 

\begin{figure}[ht]
\includegraphics[width=0.4\textwidth] {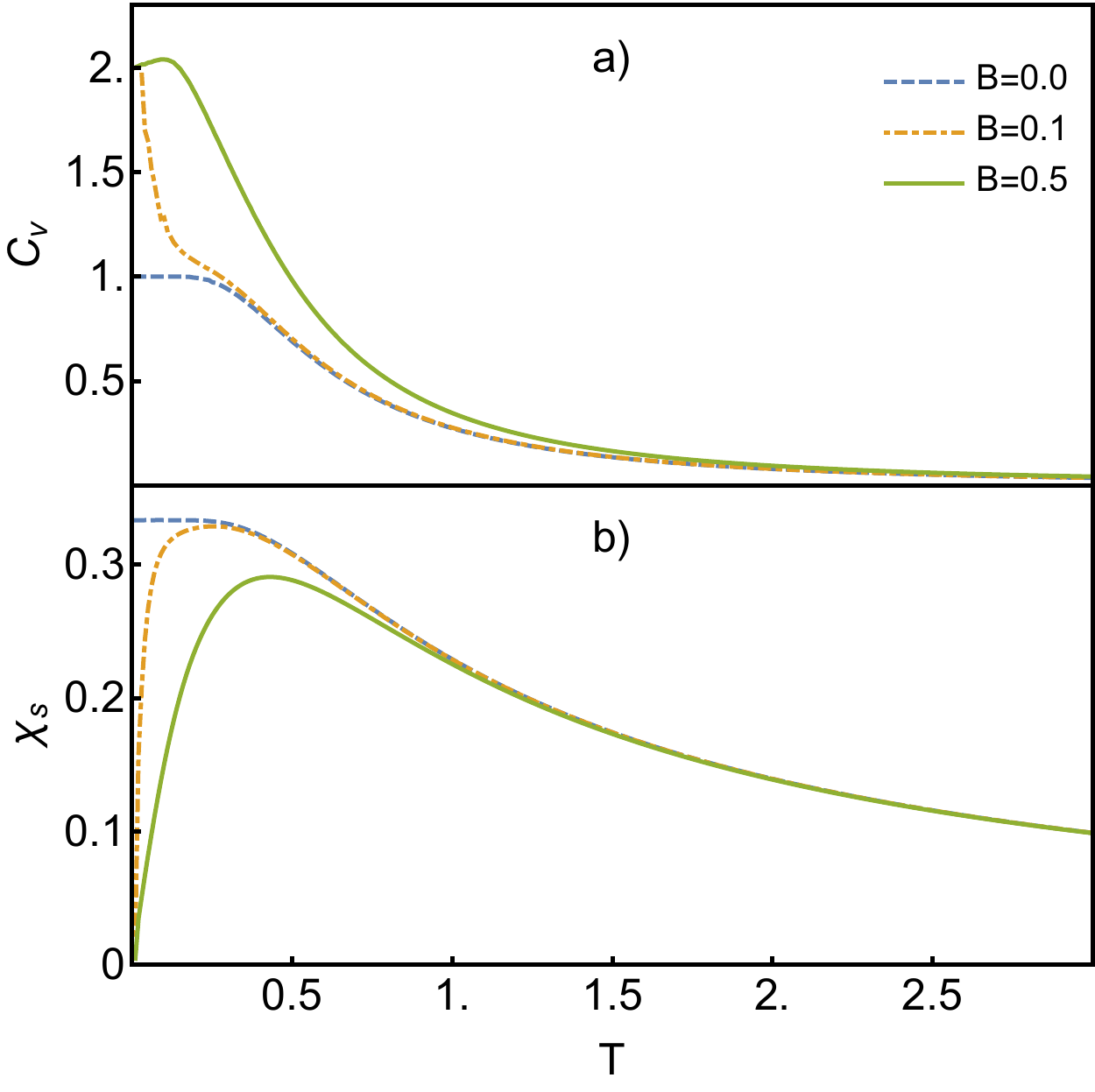}
\caption{(Color online) Specific heat $C_V$ and uniform susceptibility $\chi_s$ vs. $T$ for the pure case $\delta J=0$
and different values of the staggered field $B$.}
\label{fig3}
\end{figure}
In Fig.~3 we first present $C_V(T)$ as well as $\chi_s(T)$,  needed to evaluate  $l_t$ and $l_s$, respectively.
Their  dependence on disorder $\delta J$ (not presented) is quite weak. More significant and important for further 
discussion is  the dependence on the staggered field $B$.
We note that in contrast to the well known $S=1/2$ quantum case at $B=0$  the classical model has a 
different low-$T$ dependence for $C_V \sim k_B$  at $T \ll J$, while for $T > J$ one gets 
$C_V \sim 1/(3T^2)$. On the other hand, the behavior of $\chi_s(T)$ has at $B=0$ a 
similarity to the 1D $S=1/2$ AFM, approaching a constant for $T \to 0$. 

\begin{figure}[ht]
\includegraphics[width=0.4\textwidth] {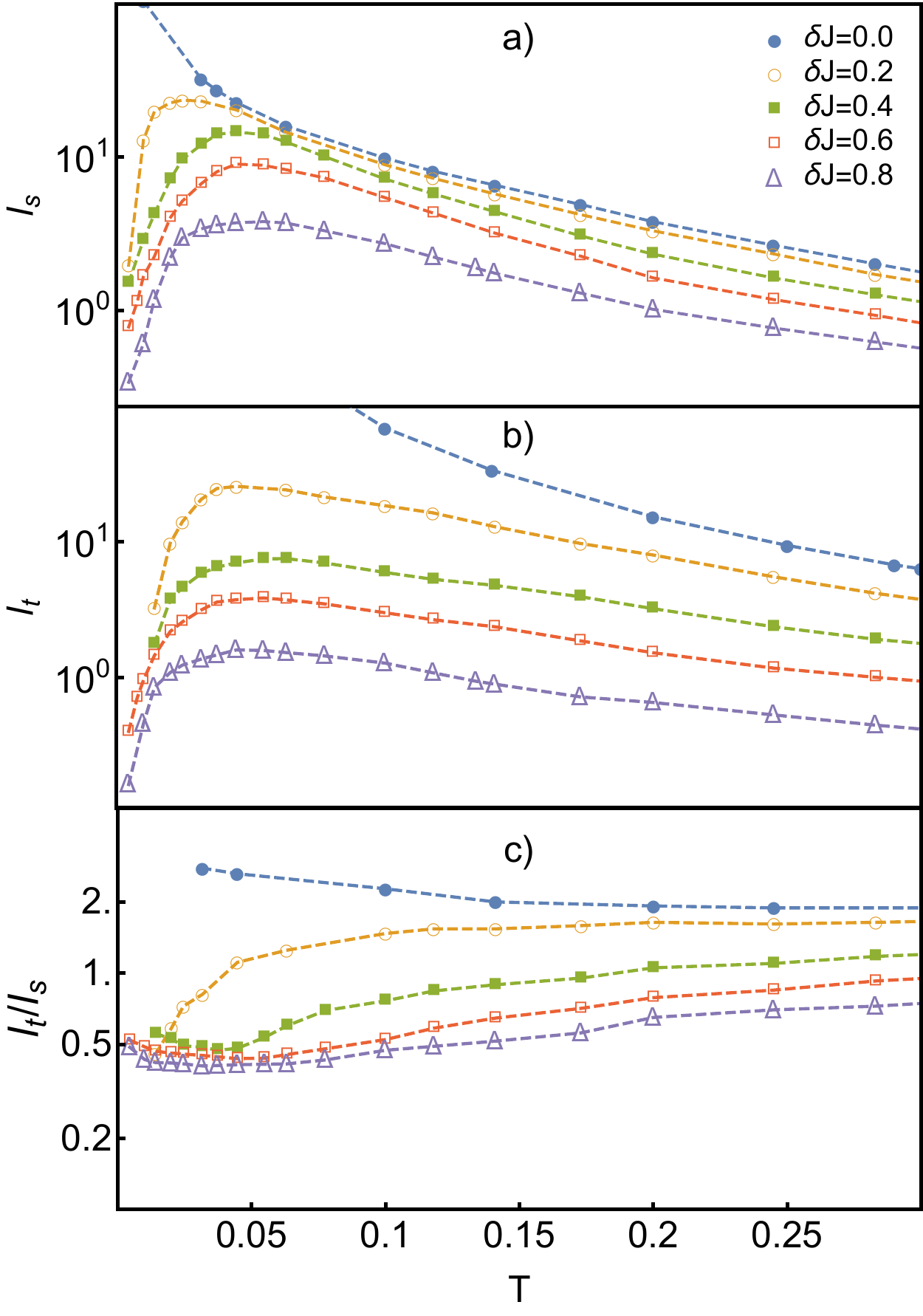}
\caption{(Color online) 
{Transport mean free paths vs. $T$: 
a)  spin diffusion $l_s(T)$, b) the thermal $l_t(T)$, and c) the Wiedemann-Franz ratio 
$W=l_t/l_s$ for different disorder  strength $\delta J=0 - 0.8$
and $B=0$. }
Note the  log scale.}
\label{fig4}
\end{figure}
In Fig.~4 we show (note the log scale) the extracted $T$ dependence of the spin and thermal MFP $l_{s,t}$,
respectively,  for $B=0$ but different disorder $0 \geq \delta J < 1$.   The qualitative and even quantitative 
behavior  of both quantities is quite similar which is not surprising since the equality $l_t \sim l_s$ would 
indicate the validity of the Wiedemann-Franz law \cite{ashcroft76} even for this system.  First we note that for a pure system 
$\delta J=0$ both MFP increase and diverge by lowering $T \to 0 $ which  is a normal behavior \cite{savin05}
emerging from restricted scattering processes at $T \to 0$. At higher $T>0.1$ finite disorder $\delta J >0$
reduces both MFP, but does not change the qualitative behavior. In contrast, at low $T<T^*$ 
whereby the crossover $T^*$ we define by $l_{s,t}(T^*)=\mathrm{max.}$ the Anderson localization 
mechanism sets in and the trend turns, i.e., $l_{s,t}( T \to 0) \to 0$. 
It is plausible that $T^*$ is increasing with $\delta J$, still
it is quite puzzling that the crossover appears very low,  typically at $T^* \leq 0.05 \ll 1$. 
This implies that even at $T > 0.1 $ nonlinear excitations become dominant 
over linear excitations, having analogy to AFM magnons. It should be also noted that our results 
reveal (at $B=0$) even at modest $\delta J>0.4$ a narrow but numerically well resolved regime
$T< 0.01 < T^*$, where both MFP become smaller than the lattice spacing $l_{s,t} <1$. 

In order to enhance the localization regime, we consider further 
the influence of the staggered magnetic field $B \neq 0$. Such a field has
a great impact on chain properties, e.g., in the quantum case this would lead to a spin
gap in the excitation spectrum. As evident from Fig.~3 $B \neq 0$ has an to some extent 
analogous effect also in the classical chain, in particular  $\chi_s(T \to 0) \to 0$ 
as well well as $C_V(T \to 0) \to 2$ for $B \neq 0$.  But the most dramatic effect is on the transport 
properties. In Fig.~5 we present results for $l_{s,t}(T)$ obtained for $B=0 - 0.5$ 
at fixed disorder $\delta J =0.4$. It is evident that  $B>0$ enhances the 
localization crossover temperature $T^*$  to values $T^* \propto B$. Moreover, the
values of $l_{s,t}$ are strongly reduced at low $T<T^*$ even at modest $\delta J$ 
and small $B=0.1$ where the maximum  at $T=T^*$ hardly goes beyond the 
'minimum metallic one', i.e. $l_{s,t} > 1$. In $B>0.1$ the transport MFP are clearly below 
the naive localization  criterion at any $T>0$.  
The origin of the localization regime enhanced by $B>0$ is not hard to rationalize,
since the introduction of $B \neq 0$ extends the validity of the linear
approximation to Eqs.~(\ref{eqm}) to much higher  $T \propto B$ and consequently also 
to the enhanced  manifestation of the Anderson localization. 
\begin{figure}[ht]
\includegraphics[width=0.4\textwidth] {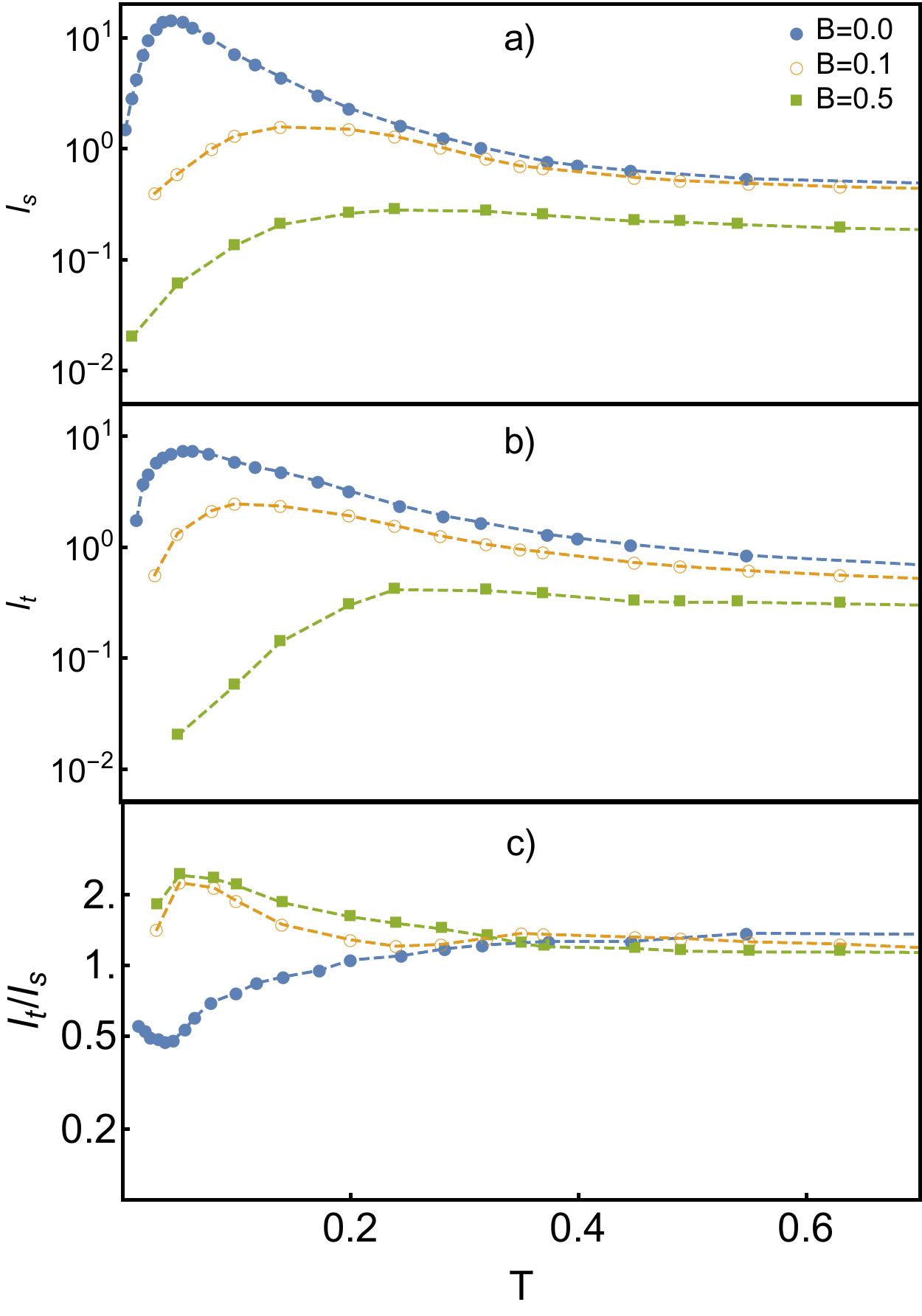}
\caption{(Color online) Transport mean free paths vs. $T$:  a) 
spin diffusion $l_s(T)$, b) thermal $l_t(T)$, and c) the ratio  $W=l_t/l_s$ for  different 
staggered fields $B=0 - 0.5$ and fixed disorder  $\delta J=0.4$}
\label{fig5}
\end{figure}

In connection to results in Figs.~4,5 it is also interesting to comment on the relation  
between both MFP: $l_s(T)$ and $l_t(T)$. Well known Wiedemann-Franz law
\cite{ashcroft76}, mostly valid for transport in metals,  translated to our system
would require an equality of both MFP $l_t \sim l_s$ in the low (i.e. $T<J$) regime.
Indeed we find that in the whole range of $T$ presented in Fig.~5 the ratio
$W=l_t/l_s$  ranges $0.5<W<2$ in the whole presented $T$ regime, and this
in spite of wide range of actual values of $10^{-2}<l_{s,t}<10 $. This implies that
the concept of constant W makes sense not just at modest scattering but even in 
the quasi-localized (nearly-localized) regime. 

\section{Conclusions}

In conclusion, presented numerical results for the spin and thermal transport in classical
disordered spin chain reveal some similarities as well as differences to the
transport in quantum spin chains. First of all, numerical studies for classical spin 
models can be performed for large systems reaching reliable results for transport quantities
even for low $T \ll J$ and arbitrary disorder which remains a severe challenge for quantum $S=1/2$
models \cite{karahalios09}, even more for larger $1/2 < S < \infty$. 

In the pure case $\delta J=0$ the $S=1/2$  model is specific due to the
integrability and consequently the ballistic transport \cite{zotos04}. 
Hence only the disorder can  serve as the scattering mechanism \cite{karahalios09,barisic10}. 
On the other hand, in the classical model without disorder $\delta J=0$
the mean-free paths remain finite for any $T>0$, with diverging $L_t,l_s \to \infty$ for $T \to 0$,
and in this sense qualitatively approaching  the dissipationless  transport in pure $S=1/2$ models 
\cite{zotos99,klumper02}. Still it should be reminded that there are qualitative difference between 
$S=1/2$ and classical spin model at $T \to 0$ in static quantities, in particular in the specific heat 
$C_V(T)$.

The most interesting common aspect of 1D classical and quantum spin $S=1/2$ transport is the 
onset of localization due to disorder $\delta J>0$ in both models at $T \to 0$. 
It is evident from our results, that there exists for any $\delta J>0$ a characteristic $T^*$,
where for $T<T^*$ transport MFP $l_t,l_s$ decrease and eventually vanish with $T \to 0$,
which can be understood with the dominant role of the linearized equations of motion, Eqs~(\ref{eqm}),
and the Anderson localizastion of the eigen-solution. On the other hand, for $T>T^*$ the transport  
is already dominated by nonlinearity effects. Our study shows
that within a classical model there is no abrupt transition between both regimes but rather
a crossover at $T \sim T^*$ where mean-free paths $l_{s,t}$ are both maximum. It is
remarkable that in a isotropic model the crossover, which indicates the  onset of nonlinear 
excitations, appears already at very low $T^* \ll J$.  Such a low-$T$ regime could
be hardly reached in numerical analysis of quantum models. 
On the other hand, our results confirm the finding in the quantum $S=1/2$ case that
the introduction of local fields $B_i$, both staggered as well as the random ones, can 
induce even stronger effect on the transport properties \cite{karahalios09,barisic10}.
This brings our model and results closer the presumable many-body localization \cite{basko06,
oganesyan07,bandarson12} but still as an approximate description.

It should be also noted that our results at lower $T<J$ agree with the qualitative validity of 
the Wiedemann-Franz law, i.e. equality of MFP $l_t \sim l_s$, recovered even in nearly 
localised regime $l_{s,t} \ll 1$.

This work has been supported by the Programs P1-0044 and P1-0112 
of the Slovenian  Research Agency (ARRS).

\end{document}